Ms. No.: JSB-13-122

**Three-dimensional network of Drosophila brain hemisphere**

Ryuta Mizutani [a,]*, Rino Saiga [a], Akihisa Takeuchi [b], Kentaro Uesugi [b], and Yoshio Suzuki [b]

[a] Department of Applied Biochemistry, School of Engineering, Tokai University, Kitakaname, Hiratsuka, Kanagawa 259-1292, Japan

[b] Research and Utilization Division, JASRI/SPring-8, Kouto, Sayo, Hyogo 679-5198, Japan

*Corresponding author.

E-mail address: ryuta@tokai-u.jp

Postal address: Department of Applied Biochemistry, School of Engineering, Tokai University, Kitakaname, Hiratsuka, Kanagawa 259-1292, Japan.

Phone: +81-463-58-1211 ext. 4184.

**Abstract**

The first step to understanding brain function is to determine the brain's network structure. We report a three-dimensional analysis of the brain network of the fruit fly *Drosophila melanogaster* by synchrotron-radiation tomographic microscopy. A skeletonized wire model of the left half of the brain network was built by tracing the three-dimensional distribution of X-ray absorption coefficients. The obtained models of neuronal processes were classified into groups on the basis of their three-dimensional structures. These classified groups correspond to neuronal tracts that send long-range projections or repeated structures of the optic lobe. The skeletonized model is also composed of neuronal processes that could not be classified into the groups. The distribution of these unclassified structures correlates with the distribution of contacts between neuronal processes. This suggests that neurons that cannot be classified into typical structures should play important roles in brain functions. The quantitative description of the brain network provides a basis for structural and statistical analyses of the Drosophila brain. The challenge is to establish a methodology for reconstructing the brain network in a higher-resolution image, leading to a comprehensive understanding of the brain structure.

**Key words:** Microtomography, micro-CT, brain, network, model building.

## 1. Introduction

A brain is composed of a huge number of neurons, which constitute an interconnected three-dimensional network. The first step to understanding brain function is to determine the brain's network structure. Several methods for visualizing the network structure have been proposed (Lichtman and Denk, 2011). Serial sectioning and subsequent light- and electron-microscopic analyses have been used as a brain tissue imaging method (Micheva and Smith, 2007; Knott *et al.*, 2008; Cardona *et al.*, 2010; Li *et al.*, 2010; Briggman *et al.*, 2011). Transgenic strategies for imaging a brain network by labeling neurons with multiple fluorescent colors have been reported (Livet *et al.*, 2007; Hampel *et al.*, 2011). Reconstruction of a virtual fly brain has been performed by labeling single neurons and superposing them (Chiang *et al.*, 2011; Peng *et al.*, 2011).

Although three-dimensional images of brain tissues have been illustrated with these methods, the network structures are difficult to comprehend. This is because of a lack of quantitative and integrative description of the three-dimensional microstructure, which should be represented in terms of three-dimensional Cartesian coordinates, rather than a three-dimensional distribution of image intensities. In molecular biology, the quantitative description of a three-dimensional molecular structure has been accomplished by building a skeletonized wire model in an electron density map obtained by crystallography (Jones and Kjeldgaard, 1997; McRee, 1999; Morris *et* *al*., 2003; Emsley and Cowtan, 2004; Steiner and Rupp, 2012). Although it is practically impossible to comprehend the whole picture of the electron density map itself, step-by-step building of a wire model can be performed by placing atoms in the map. The resultant model is regarded as a molecular structure that provides us with some understanding of the mechanisms of molecular recognition and biochemical reaction (Blundell and Johnson, 1976). A similar procedure can be applied to the three-dimensional analysis of brain networks. We have reported skeletonized wire models of human brain networks reconstructed in Cartesian coordinate space (Mizutani *et al*., 2010). Structures of a selected subset of mouse retinal cells have been manually reconstructed from image stacks obtained by serial-sectioning electron microscopy (Briggman *et al*., 2011). The skeletonization facilitates our understanding of these network structures composed of a number of constituents.

In this paper, we report a three-dimensional analysis of the brain network of the fruit fly *Drosophila melanogaster*. The fly brain, also referred to as the cephalic ganglion, was labeled with aurate dye by reduced-silver impregnation and then visualized by X-ray tomographic microscopy, which is known as micro-CT or microtomography (Bonse and Busch, 1996; Mizutani and Suzuki, 2012). By subjecting the tomographic image to model building, we obtained a three-dimensional structural description of the left half of the fly brain. This quantitative description provides a basis for structural and statistical analysis of the brain network.

## 2. Material and Methods

*Sample preparation*

Wild-type fruit flies *Drosophila melanogaster* Canton-S (Drosophila Genetic Resource Center, Kyoto Institute of Technology, Japan) were raised on standard cornmeal-molasses fly food and kept at 20°C. Adult fly brains were dissected and fixed for 30 min in a solution containing 85% ethanol, 5% acetic acid, and 3.8% formaldehyde. After being washed with ethanol, the sample was progressively hydrated in 90%, 70%, 50%, and 30% ethanol and then in distilled water. The hydrated sample was subjected to modified reduced-silver impregnation (Heisenberg, 1989; Mizutani *et al.*, 2007). The impregnation was performed for 15 hr at 37°C using 0.038% silver nitrate solution containing 0.75% pyridine and 77 mM sodium borate (pH 8.4). The sample was then developed for 10 min at 25°C in a solution containing 1% hydroquinone and 10% sodium sulfite. After being washed with distilled water twice, the sample was toned with 1% hydrogen tetrachloroaurate for 60 min. After being washed again, the sample was immersed sequentially in 2% oxalic acid for 10 min and 5% sodium thiosulfate for 5 min, with two 2-min washes in distilled water between immersions. After removal of the residual sodium thiosulfate by washing, the sample was dehydrated in an ethanol series and transferred to n-butyl glycidyl ether. The sample was then equilibrated with Petropoxy 154 epoxy resin (Burnham Petrographics, ID, USA) at 4°C for 24 hr. The epoxy resin was degassed in vacuum before use. The sample was again placed into a new resin aliquot and equilibrated for an additional 24 hr at 4°C. The obtained sample was mounted using a nylon loop (Hampton Research, CA, USA) and then incubated at 90ºC for 16 hr to cure the epoxy resin. Paraffin sections for light microscopy were also prepared from brains stained with the same procedure.

*Tomographic microscopy*

Three-dimensional distributions of X-ray absorption coefficients were determined by tomographic microscopy at the BL47XU beamline of the synchrotron radiation facility SPring-8, as described previously (Takeuchi *et al*., 2006). The sample was mounted on the microtomograph by using a brass fitting designed for biological samples. A Fresnel zone plate with outermost zone width of 100 nm and diameter of 774 μm was used as an X-ray objective lens to attain a viewing field wider than the sample width. Two-dimensional transmission radiographs produced by 8-keV X-rays were recorded using a CCD-based X-ray imaging detector (AA40P and C4880-41S, Hamamatsu Photonics, Japan). The viewing field was 2000 pixels horizontally × 700 pixels vertically. The effective pixel size was 262 nm × 262 nm. In total, 1800 images per dataset were acquired with a rotation step of 0.10º and exposure time of 300 ms. The data collection conditions are summarized in Table 1.

The obtained radiographs were subjected to convolution-back-projection calculation using the program RecView (Mizutani *et al*., 2011; available from http://www.el.u-tokai.ac.jp/ryuta/) accelerated with CUDA parallel-computing processors. In total, 700 tomographic slices perpendicular to the sample rotation axis were reconstructed from one dataset with this calculation, giving linear absorption coefficients at 8 keV in a cylindrical region with a diameter of 2000 voxels and height of 700 voxels. The spatial resolution of the reconstructed image was estimated from values of the modulation transfer function by using three-dimensional square-wave patterns (Mizutani and Suzuki, 2012).

Since the sample dimension of 570 μm along the sample rotation axis was larger than the viewing field height, four datasets were taken by displacing the sample 150 μm along the rotation axis so as to cover the entire sample. To determine the positional relationship of these datasets, each end of the three-dimensional image was superposed by minimizing the root-mean-square difference in the linear absorption coefficients by using the program RecView. The whole image was built up by stacking these datasets. Finally, regions corresponding to the outer resin and the nylon loop used for sample mounting were trimmed to obtain a three-dimensional image within a rectangular box nearly circumscribing the brain. The data size of the obtained image composed of

840 × 1250 × 2176 voxels was 2.3 GB.

*Model building*

A three-dimensional region of 840 × 1250 × 1200 voxels of the obtained image covering the left half of the brain was subjected to the model building procedure by using the program MCTrace (Mizutani et al., 2011; available from http://www.el.u-tokai.ac.jp/ryuta/). Neuronal processes were traced by tracking the absorption coefficient distribution down to 19.4 cm$^{-1}$. This is equivalent to 1.7 times the standard deviation (1.7 σ) of absorption coefficients in the volume subjected to the model building. The model building was performed using a method like those reported for crystallographic studies of macromolecular structures (Jones and Kjeldgaard, 1997; McRee, 1999; Morris et al., 2003; Emsley and Cowtan, 2004; Steiner and Rupp, 2012). Node coordinates of a congested neuropil were placed and connected while the three-dimensional map of absorption coefficients was viewed. Sparsely distributed structures were built by automatic tracing (Mizutani et al., 2011) and examined manually afterward. To accelerate these building procedures, functions specially designed for model manipulation, such as edit functions that require minimum manual interventions and comment functions for attaching comments to model nodes, were implemented. The model building statistics are summarized in Table 1.

A set of nodes linked through connections was designated as a "trace" and numbered serially.

Each node belonging to a certain trace was discriminated using node names. Traces were classified into groups according to the three-dimensional structures. Therefore, each node was designated by node name, trace number, and group name. Traces that could not be classified into any groups were left unclassified.

The coordinate system of the model was defined so as to place the +x/-x axis nearly along the anteroposterior direction, the +y/-y axis along the dorsoventral direction, and the +z/-z axis along the right-to-left direction. The coordinate origin was placed at the origin of the coefficient map. Cartesian coordinates of model nodes given in micrometers were enumerated in Protein Data Bank format and posted in the microtomographic structure repository on the institution's web page (http://www.el.u-tokai.ac.jp/ryuta/).

*Structure analysis*

The obtained network model is composed of sequential line segments connecting Cartesian coordinates of the model nodes. Positional relationships between one terminus of a certain neuronal process and the other part of the model were estimated by calculating the minimum distance from that terminus to the other line segments composing the model. Though we have not established a method for estimating coordinate errors of the microtomographic model, the spatial resolution should be taken into account in evaluating the positional relationships. In this study, neuronal process pairs with distances less than 1.2 μm (corresponding to 1.5 times the spatial resolution) were regarded as candidates for neuronal connections and designated as having "contacts". Contacts within each classified group were not included since constituents of a certain group exhibited similar structures.

The volume occupied by the model was estimated using the model envelope composed of spheres and cylinders, as reported previously (Mizutani *et al.*, 2011). The volume subjected to the model building was estimated by counting the number of voxels that lay inside the circumscribing polygon of the model in each tomographic slice, and exhibited absorption coefficients greater than 1.7 σ (19.4 cm$^{-1}$) or 3.0 σ (34.3 cm$^{-1}$) of the absorption coefficients. Model coordinates of the peripheral nerves were excluded from the volume estimation.

*High-resolution microtomographic analysis*

High-resolution microtomographic visualization was performed at the BL47XU beamline of SPring-8, as described previously (Takeuchi *et al.*, 2009). A Fresnel zone plate with outermost zone width of 50 nm and diameter of 155 μm was used as an X-ray objective lens. Zernike phase-contrast radiographs produced by 8-keV X-rays were recorded using a CCD-based X-ray imaging detector (AA40P and C4880-41S, Hamamatsu Photonics, Japan). The viewing field was 1136 pixels horizontally × 940 pixels vertically. The effective pixel size was 21.8 nm × 21.8 nm. 450 images per dataset were acquired with a rotation step of 0.40º and exposure time of 200 ms.

## 3. Results and Discussion

*Three-dimensional visualization*

A rendering of the entire three-dimensional distribution of X-ray absorption coefficients of the Drosophila brain stained with the reduced-silver impregnation is shown in Fig. 1A. The spatial resolution was estimated by using a microfabricated test object (Mizutani and Suzuki, 2012). The resolution was 600 nm along the sample rotation axis, which is nearly parallel to the right-to-left direction of the brain, and 800 nm along the direction perpendicular to this rotation axis. The Drosophila brain exhibits a bilaterally symmetrical structure composed of left and right optic lobes and the central brain (Strausfeld, 1976), as shown in Fig. 1A and Supplementary video 1. Data collection conditions are summarized in Table 1.

Fig. 2A shows a magnified light-microscopy image of a horizontal section of a Drosophila brain stained with the same procedure. Neuronal processes of the fan-shaped body observed in this 3-μm paraffin section exhibit structures corresponding to small field neurons of the central complex (Young and Armstrong, 2010). Neuronal processes of this region were reproduced in a corresponding cross section of the three-dimensional distribution of X-ray absorption coefficients (Fig. 2B). The neuronal processes exhibited approximate diameters of 0.5–1.0 μm in the three-dimensional coefficient distribution, suggesting that the spatial resolution determined using the microfabricated test object has also been achieved in the brain image.

*Building skeletons of neuronal processes*

The absorption coefficient distribution was traced to build skeletonized wire models of the neuronal processes. In the model building procedure, three-dimensional coordinates were chosen along continuous coefficient distributions to trace neuronal processes. The three-dimensional coefficient map of the fan-shaped body superposed on the skeletonized models is shown in Fig. 1B. Through this model building procedure, neuronal process structures observed in the absorption coefficient distribution (Fig. 2B) were represented in three-dimensional coordinates, giving the skeletonized models (Fig. 2C).

Model building statistics are summarized in Table 1. Since bushy structures of terminal arbors were not entirely resolved at this resolution, their coefficient distributions were traced as simplified models, as has been performed in the skeletonization of fluorescence images of single neurons (Chiang *et al.*, 2011; Lai *et al.*, 2012). The reduced silver impregnation exhibited fragmented structures for some of neuronal processes. These fragmented neuronal processes were built as individual traces. Most of the cell bodies were not built, mainly because of the lack of pairwise connections between neuropil and cell bodies.

The number of neurons in the adult Drosophila brain has been estimated (Shimada *et al*., 2005) to be between 95,000 and 110,000. In this study, fascicles of neuronal processes that could not be resolved in the coefficient map were built as unified structures. This resulted in a reduced number of traces, corresponding to one third of the estimated number of neurons in the brain half. The obtained model occupied 83% of the volume of the traced region with absorption coefficients greater than 3 σ, or 47% of the volume for those greater than 1.7 σ, though some tissue structures such as mushroom body were only partly visualized by the reduced-silver impregnation method (Strausfeld, 1976) used in this study.

The entire model is shown in Fig. 2C. Although we have reported automatic model building for the coefficient maps of human cerebral cortex (Mizutani *et al*., 2010), this method was efficient only for the skeletonization of sparsely distributed structures, but not for the congested neuropil of fly brain. Therefore, in this study, automatic tracing was applied to sparsely distributed structures such as those of peripheral nerves, while manual tracing was used to build the neuropil model. It took approximately 1700 person-hours (an average of 150 coordinate node incorporations into the model per person-hour) of this combination of automatic and manual model building to obtain a skeletonized model with a total length of 378 mm. The performance of model building was 4.5 hours per mm of processes. This is comparable to the human workload of 5.3 hours per mm reported for building three-dimensional models from serial-sectioning electron microscopy images (Helmstaedter *et al*., 2011). Several methods for automatically tracing a brain network have also been proposed (Liu, 2011). Improvements in these tracing methods should facilitate accurate interpretation of three-dimensional images.

The obtained models of neuronal processes were classified into 362 groups on the basis of their three-dimensional structures. The classified groups were designated with three letters and numbers beginning with N (left half) or M (right half). These classified groups represent neuronal processes that send long-range projections or repeated structures of the optic lobe. Neuronal processes that could not be classified into any groups were observed mostly in the central brain.

*Classified structures of the optic lobe*

The optic lobe structure is shown in Fig. 3 and Supplementary video 2. The medulla of the optic lobe is composed of two structural categories of groups: groups nearly perpendicular to the outer surface of the brain tissue and groups parallel to the brain surface (Fig. 3B). The groups perpendicular to the surface (NC3–NCZ) exhibit structures corresponding to Mi and Tm neurons of the medulla (Fischbach and Dittrich; 1989). The groups parallel to the surface (NF1–NF5,

NF8, NF9) correspond to structures of Mt neurons (Fischbach and Dittrich, 1989) and pass posteriorly as the posterior optic tract.

The second optic chiasma exhibits a converging Y-shaped structure, as shown in Fig. 3A. The wide openings of these Y-shaped structures (NB1–NBR) are oriented toward the medulla. The Y-shaped stems are pointed posteriorly and are structurally divisible into 25–27 segments (Fig. 3C). In spans of these stems, neuronal processes of the lobula (ND1–NDI) are interlaced approximately orthogonal to the stems (Fig. 3C). These structures correspond to those of Lt neurons (Fischbach and Dittrich, 1989). A cage structure is formed in the lobula region proximal to the central brain (Fig. 3D). The constituents of this cage structure were classified into distal (NDW), intermediate (NDV) and proximal (NDU, NDX, NDZ) layers.

The structure of the lobula plate is shown in Fig. 3E. A posterior face of the lobula plate (NBY) corresponds to structures of vertical motion sensitive (VS) neurons (Scott *et al*., 2002). An anterior face adjoining the second optic chiasma (NBX and NBZ) corresponds to structures of horizontal motion sensitive (HS) neurons (Schnell *et al*., 2010). The NBW group forms an intermediate stratum between these faces of the lobula plate and passes contralaterally to the right half brain through the sub-ellipsoid commissure.

*Classified structures of the central brain*

The central brain exhibits irreducible structures in contrast to the repeated structures of the optic lobe. The structures of the central brain network involving neuronal processes from the lobula are shown in Fig. 4A. One of the neuronal process groups from the lobula (NH8) constitutes the anterior optic tract and exhibits a three-dimensional structure corresponding to LC6, LC9, and LC10 neurons (Otsuna and Ito, 2006). These networks associated with the lobula cover a number of brain regions, indicating that widespread transmission of visual information can be performed with these structures.

The network of the fan-shaped body, ellipsoid body, and noduli is shown in Fig. 4B. Groups composing most of the fan-shaped body (NM2 and NM4) exhibit structures corresponding to small field neurons of the central complex (Young and Armstrong, 2010). These groups are located in the vicinity of the posterior ends of the NJ2 and NJ3 groups, which are distributed to the anterior optic tubercle as shown in Fig. 4A. It has been suggested that the fan-shaped body is involved in a number of functions, including motor behavior control and visual memory storage (Strauss, 2002; Liu *et al*., 2006). The neuronal processes of these NJ2 and NJ3 groups can transmit visual information required for the high-order functions achieved by the fan-shaped body.

The antennal lobe is known as the first processing area of olfactory information (Masse *et al*., 2009). The network involving the antennal lobe is shown in Fig. 4C. The structure of the antennal lobe is composed of neuronal processes distributed locally (NKI, NKJ, NL5, NL9, and NLA) and those distributed widely to other brain regions (NI6, NI8, NIB, NIZ, and NN7). These structures coincide with the organization of the antennal lobe composed of local neurons and projection neurons (Masse *et al*., 2009). Some neuronal processes (NIB and NN7) pass contralaterally to the right half brain through the antennal commissure (Stocker *et al*., 1990). The network involving the antennal lobe is also associated with neuronal processes of outer (NHB), middle (NIY), and inner (NHG) antennocerebral tracts. These antennocerebral tracts distribute to the lateral horn, which is known as a higher olfactory center (Masse *et al*., 2009).

*Unclassified structures of the central brain*

The skeletonized model is also composed of neuronal processes that could not be classified into the groups. These unclassified structures composed of 4323 traces exhibited a wide distribution over the central brain. The unclassified structures should represent neurons that innervate locally or that exhibit unique structures. Although most of the neuronal structures have been described by classifying them into groups (Fischbach and Dittrich, 1989; Scott *et al*., 2002; Otsuna and Ito, 2006; Jefferis *et al*., 2007; Schnell *et al*., 2010), little is known about neurons that cannot be classified into typical structures. Statistical analysis can also be applied to such unclassified structures. Fig. 5 shows the spatial distribution of contacts in frontal sections of the skeletonized model. The observed contacts showed co-localization with the unclassified structures rather than with the classified groups. Although synaptic connections at these contacts cannot be visualized with the spatial resolution achieved in this study, neuronal processes should be located in close proximity to form synapses. Therefore, the contacts shown in Fig. 5 should partly represent the distribution of neuronal connections.

In Fig. 6, the number of contacts in a unit volume of $20 \times 20 \times 20$ μm$^3$ is plotted against the total length of unclassified or classified traces in that unit volume. The number of contacts showed a positive correlation with the length of the unclassified traces (Fig. 6), but little correlation with the length of the classified traces. This indicates that the unclassified structures have contacts depending on their length, while the classified structures have contacts that do not depend on the length. The length dependence of the unclassified traces was not observed when contacts within the unclassified structures were excluded from the analysis. Therefore, the number of contacts involving the unclassified structures should correlate with the degree of congestion of the unclassified traces, which can be represented with the length of the unclassified traces per volume. These results suggest that neurons that cannot be classified into structural groups should play important roles in brain functions, though their structures have hardly been investigated.

## 4. Conclusion

Although it is difficult to comprehend the three-dimensional distribution of image intensities itself, the skeletonized model built by tracing the distribution of X-ray absorption coefficients allowed quantitative analyses of the network architecture. The statistical analysis of three-dimensional coordinates of the network structure revealed a correlation between the number of contacts and the length of the unclassified traces per volume. However, there still remain difficulties in figuring out the total of over 15,000 traces comprising the model. The structural classification reduced a number of structural units by compiling neuronal processes into a limited number of groups. Structures of interest such as those of the optic lobe can be extracted from the entire model by specifying neuronal processes in a group-by-group manner. Handling the extracted subset of the three-dimensional structure is easier than handling every trace. Therefore, the quantitative description of the network structure along with the structural classification can improve the accessibility of the network structure for even those who are not familiar with the brain architecture.

The spatial resolution of microscopy is primarily constrained by the diffraction limit of the observing wavelength. Since the wavelength of X-rays is much shorter than that of visible light, the diffraction-limited resolution of X-ray microscopy is much higher than that of conventional light microscopy. Instead, the three-dimensional spatial resolution of X-ray microtomography is constrained by the specifications of the X-ray optics used for radiographic imaging. In this study, a Fresnel zone plate with outermost zone width of 100 nm was used as an X-ray objective lens to acquire radiographs. Although the spatial resolution achieved with these X-ray optics was mesoscopic and estimated to be 600–800 nm, the three-dimensional image reconstructed from the datasets had a total file size of 2.3 GB. Spatial resolution itself can be pursued to resolve much finer structures. Fig. 7 shows a sagittal view of a high-resolution microtomographic image of the same brain sample. The spatial resolution of this analysis was estimated to be 140 nm by using a microfabricated test object. Neuronal processes can further be resolved at this spatial resolution. However, the 5-fold improvement in the spatial resolution leads to a 125-fold increase in the data size of the three-dimensional image. In this study, it took 1700 person-hours to build the skeletonized model. It will take many more person-hours to build a model in this high-resolution image. Therefore, at least in this respect, the reconstruction of brain network in the high-resolution image appears to be prohibitively expensive in terms of human workload (Helmstaedter *et al*., 2009).

However, the first step toward an understanding of the brain network is to map its elements and connections comprehensively in order to create a structural description of the network architecture (Sporns *et al*., 2011). Although three-dimensional images of brain networks have been reported, only subsets of network constituents have been reconstructed. This is because of the difficulties in reconstructing brain networks quantitatively and comprehensively in high-resolution three-dimensional images. In this study, manual model building was combined with automatic tracing to reconstruct the brain network in Cartesian coordinate space. Computed interpretation of neuronal process structures (Liu, 2011) should allow automatic model building in a three-dimensional image with higher spatial resolution. The challenge is to establish a methodology for this time-consuming step, which accounts for most of the brain structure analysis.

**Acknowledgments**

We thank Yasuo Miyamoto and Kiyoshi Hiraga (Technical Service Coordination Office, Tokai University) for helpful assistance with the microfabrication of the test object and preparation of the brass fittings for tomographic data collection. We thank Noboru Kawabe (Teaching and Research Support Center, Tokai University School of Medicine) for helpful assistance with tissue sample preparation. This work was supported in part by a Grant-in-Aid for Scientific Research from the Japan Society for the Promotion of Science (nos. 21611009, 25282250, and 25610126). The synchrotron radiation experiments were performed at SPring-8 with the approval of the Japan Synchrotron Radiation Research Institute (JASRI) (proposal nos. 2007B1102, 2008A1190, 2008B1261, 2011B0041, and 2012B0041).

**References**

Blundell, T.L., Johnson, L.N., 1976. Protein Crystallography. Academic Press, London.

Bonse, U., Busch, F., 1996. X-ray computed microtomography (microCT) using synchrotron radiation (SR). Prog. Biophys. Mol. Biol. 65, 133-169.

Briggman, K.L., Helmstaedter, M., Denk, W., 2011. Wiring specificity in the direction-selectivity circuit of the retina. Nature 471, 183-188.

Cardona, A., Saalfeld, S., Preibisch, S., Schmid, B., Cheng, A., et al., 2010. An integrated micro- and macroarchitectural analysis of the Drosophila brain by computer-assisted serial section electron microscopy. PLoS Biol. 8, e1000502.

Chiang, A.S., Lin, C.Y., Chuang, C.C., Chang, H.M., Hsieh, C.H., et al., 2011. Three-dimensional reconstruction of brain-wide wiring networks in Drosophila at single-cell resolution. Curr. Biol. 21, 1-11.

Emsley, P., Cowtan, K., 2004. Coot: model-building tools for molecular graphics. Acta Crystallogr. D60, 2126-2132.

Fischbach, K.-F., Dittrich, A.P.M., 1989. The optic lobe of Drosophila melanogaster. I. A Golgi analysis of wild-type structure. Cell Tissue Res. 258, 441-475.

Hampel, S., Chung, P., McKellar, C.E., Hall, D., Looger, L.L., et al., 2011. Drosophila Brainbow:

a recombinase-based fluorescence labeling technique to subdivide neural expression patterns. Nat. Methods 8, 253-259.

Heisenberg, M., 1989. Silver staining the central nervous system, in: Ashburner, M. (Eds), Drosophila: a Laboratory Manual. Cold Spring Harbor Laboratory Press, New York, pp. 277-278.

Helmstaedter, M., Briggman, K.L., Denk, W., 2009. 3D structural imaging of the brain with photons and electrons. Curr. Opin. Neurobiol. 18, 633-641.

Helmstaedter, M., Briggman, K.L., Denk, W., 2011. High-accuracy neurite reconstruction for high-throughput neuroanatomy. Nat. Neurosci. 14, 1081-1088.

Jefferis, G.S., Potter, C.J., Chan, A.M., Marin, E.C., Rohlfing, et al., 2007. Comprehensive maps of Drosophila higher olfactory centers: spatially segregated fruit and pheromone representation. Cell 128, 1187-1203.

Jones, T.A., Kjeldgaard, M., 1997. Electron-density map interpretation. Meth. Enzymol. 277, 173-208.

Knott, G., Marchman, H., Wall, D., Lich, B., 2008. Serial section scanning electron microscopy of adult brain tissue using focused ion beam milling. J. Neurosci. 28, 2959-2964.

Lai, J.S., Lo, S.J., Dickson, B.J., Chiang, A.S., 2012. Auditory circuit in the Drosophila brain. Proc. Natl. Acad. Sci. U.S.A. 109, 2607-2612.

Li, A., Gong, H., Zhang, B., Wang, Q., Yan, C., et al., 2010. Micro-optical sectioning tomography to obtain a high-resolution atlas of the mouse brain. Science 330, 1404-1408.

Lichtman, J.W., Denk, W., 2011. The big and the small: challenges of imaging the brain's circuits. Science 334, 618-623.

Liu, G., Seiler, H., Wen, A., Zars, T., Ito, K., et al., 2006. Distinct memory traces for two visual features in the Drosophila brain. Nature 439, 551-556.

Liu, Y., 2011. The DIADEM and beyond. Neuroinformatics 9, 99-102.

Livet, J., Weissman, T.A., Kang, H., Draft, R.W., Lu, J., et al., 2007. Transgenic strategies for combinatorial expression of fluorescent proteins in the nervous system. Nature 450, 56-62.

Masse, N.Y., Turner, G.C., Jefferis, G.S., 2009. Olfactory information processing in Drosophila. Curr. Biol. 19, R700-713.

McRee, D.E., 1999. XtalView/Xfit--A versatile program for manipulating atomic coordinates and electron density. J. Struct. Biol. 125, 156-165.

Micheva, K.D., Smith, S.J., 2007. Array tomography: a new tool for imaging the molecular architecture and ultrastructure of neural circuits. Neuron 55, 25-36.

Mizutani, R., Takeuchi, A., Hara, T., Uesugi, K., Suzuki, Y., 2007. Computed tomography imaging of the neuronal structure of Drosophila brain. J. Synchrotron Radiat. 14, 282-287.

Mizutani, R., Takeuchi, A., Uesugi, K., Takekoshi, S., Osamura, R.Y., et al., 2010.

Microtomographic analysis of neuronal circuits of human brain. Cereb. Cortex 20, 1739-1748.

Mizutani, R., Takeuchi, A., Uesugi, K., Takekoshi, S., Nakamura, N., et al., 2011. Building human brain network in 3D coefficient map determined by X-ray microtomography. AIP Conf. Proc. 1365, 403-406.

Mizutani, R., Suzuki, Y., 2012. X-ray microtomography in biology. Micron 43, 104-115.

Morris, R.J., Perrakis, A., Lamzin, V.S., 2003. ARP/wARP and automatic interpretation of protein electron density maps. Meth. Enzymol. 374, 229-244.

Otsuna, H., Ito, K., 2006. Systematic analysis of the visual projection neurons of Drosophila melanogaster. I. Lobula-specific pathways. J. Comp. Neurol. 497, 928-958.

Peng, H., Chung, P., Long, F., Qu, L., Jenett, A., et al., 2011. BrainAligner: 3D registration atlases of Drosophila brains. Nat. Methods 8, 493-500.

Schnell, B., Joesch, M., Forstner, F., Raghu, S.V., Otsuna, H., et al., 2010. Processing of horizontal optic flow in three visual interneurons of the Drosophila brain. J. Neurophysiol. 103, 1646-1657.

Scott, E.K., Raabe, T., Luo, L., 2002. Structure of the vertical and horizontal system neurons of the lobula plate in Drosophila. J. Comp. Neurol. 454, 470-481.

Shimada, T., Kato, K., Kamikouchi, A., Ito, K., 2005. Analysis of the distribution of the brain

cells of the fruit fly by an automatic cell counting algorithm. Physica A350, 144–149.

Sporns, O., 2011. The human connectome: a complex network. Ann. N.Y. Acad. Sci. 1224, 109-125.

Steiner, R.A., Rupp, B., 2012. Model building, refinement and validation. Acta Crystallogr. D68, 325-327.

Stocker, R.F., Lienhard, M.C., Borst, A., Fischbach, K.F., 1990. Neuronal architecture of the antennal lobe in Drosophila melanogaster. Cell Tissue Res. 262, 9-34.

Strausfeld, N.J., 1976. Atlas of an Insect Brain. Springer-Verlag, Berlin.

Strauss, R., 2002. The central complex and the genetic dissection of locomotor behaviour. Curr. Opin. Neurobiol. 12, 633-638.

Takeuchi, A., Uesugi, K., Suzuki, Y., 2009. Zernike phase-contrast x-ray microscope with pseudo-Kohler illumination generated by sectored (polygon) condenser plate. J. Phys. Conf. Ser. 186, 012020.

Takeuchi, A., Uesugi, K., Suzuki, Y., Tamura, S., Kamijo, N., 2006. High-resolution X-Ray imaging microtomography with Fresnel zone plate optics at SPring-8. IPAP Conf. Series 7, 360-362.

Young, J.M., Armstrong, J.D. 2010. Structure of the adult central complex in *Drosophila*: organization of distinct neuronal subsets. J. Comp. Neurol. 518, 1500-1524.

**Figure captions**

**Fig. 1.** Structure of Drosophila brain. (A) Three-dimensional rendering of X-ray absorption coefficients of Drosophila brain visualized by tomographic microscopy. The anterior half of the brain is cut away to illustrate inner structures. Dorsal is to the top. Left and right optic lobes (L-OL, R-OL) and central brain (CB) are labeled. X-ray absorption coefficients were rendered from 15 cm$^{-1}$ (black) to 100 cm$^{-1}$ (white). Scale bar: 20 μm at the cutaway section. (B) Stereo drawing of coefficient maps superposed on a wire model (gray) of part of the fan-shaped body. Coefficient maps in a 262-nm grid are contoured at 22.9 cm$^{-1}$ (purple) corresponding to two times the standard deviation (2 σ) of absorption coefficients and at 68.6 cm$^{-1}$ (green) corresponding to 6 σ. (C) The entire network model viewed from approximately the same direction as in Fig 1A. The left optic lobe (L-OL) and central brain (CB) are labeled. Neuronal process groups are differentiated by color.

**Fig. 2.** Building skeletons of neuronal processes. (A) Light-microscopy image of a horizontal section of Drosophila brain. The brain sample was stained with the reduced-silver impregnation and embedded in paraffin to prepare 3-μm sections. The fan-shaped body (FB) and ellipsoid body (EB) are labeled. Scale bar: 10 μm. (B) Cross section of tomographic image corresponding to the paraffin section shown in Fig. 2A. X-ray absorption coefficients were rendered from 10 cm$^{-1}$ (black) to 90 cm$^{-1}$ (white). (C) Skeletonized wire models built by tracing absorption coefficients. Structures in a 20-μm slab below the cross section shown in Fig. 2B are illustrated. Neuronal process groups are differentiated by color.

**Fig. 3.** Structure of optic lobe. In Fig. 3B–3E, neuronal process groups are differentiated by color. Some groups are labeled. (A) Dorsal view of the overall structure. The medulla (Me) is drawn in green, lobula plate (LP) in orange, second optic chiasma (Ch) in blue, lobula (Lo) in purple, and other structures in gray. Anterior is to the top. (B) Structure of the medulla viewed from the direction along rows of NC3–NCZ groups. (C) Structure of the second optic chiasma and ND1–NDI groups of the lobula. In the upper panel, the structure is viewed from the brain center so as to illustrate rows of NB1–NBR groups. This orientation corresponds to a side view of Fig. 3A. In the lower panel, the structure is rotated around the horizontal axis by approximately 95° to show a view from nearly the anterior direction. (D) Structure of the lobula viewed from the direction along rows of ND1–NDI groups. The NDW group is drawn in orange, NDV in blue, and NDU, NDX, and NDZ in magenta. Dorsal is to the right. (E) Orthographic views of the lobula plate structure. The NBY group is drawn in magenta, NBX and NBZ in green, NBW in brown, and NDY in blue. In the right panel, the structure is viewed from nearly the dorsal

direction.

**Fig. 4.** Structure of central brain. Upper panels are dorsal views. Posterior is to the top. Lower panels are anterior views the same as in Fig. 1C. Dorsal is to the top. Neuronal process groups are differentiated by color. Outlines of the central brain are indicated with broken lines. Some groups are labeled. (**A**) Network associated with the lobula. Structure of the lobula (Lo) along with NH3, NH4, NH8, NHA, NHC, NHK, NJ2, and NJ3 groups are illustrated. (**B**) Structures of the fan-shaped body (FB) composed of NJ4, NJ7, NM2, NM4, and NNK groups, ellipsoid body (EB) composed of NN4, NN5, and NN6 groups, and noduli composed of NLB and NMS groups are illustrated along with NJ0 group of the superior ellipsoid commissure. (**C**) Network associated with the antennal lobe (AL). Structures of locally distributed NKI, NKJ, NL5, NL9, and NLA groups and widely distributed NI6, NI8, NIB, NIZ, and NN7 groups along with the antennocerebral tracts (NHB, NHG, and NIY) are illustrated.

**Fig. 5.** Distribution of unclassified structures. Frontal sections of the network structure viewed from the same direction as in Fig. 1C. Dorsal is to the top. Neuronal processes of classified structures within a slab of 10-μm thickness are drawn in gray, and those of unclassified structures in blue. Contacts between neuronal processes are indicated with red crosses. (A) Anterior section

around the ellipsoid body (e b). (B) Middle section around the great commissure (g c). (C) Posterior section around the posterior slope.

**Fig. 6.** The number of contacts in each volume of $20 \times 20 \times 20$ μm$^3$ is plotted against the total length of traces in the same volume. Plots against the length of classified traces are indicated with open circles and those of unclassified traces with closed circles.

**Fig. 7.** Sagittal section of central brain visualized by microtomography. Three-dimensional structure within a $90 \times 90 \times 20$ μm$^3$ region of the central brain were rendered. (A) Drosophila brain visualized at 600–800 nm resolution. Great commissure (GC) and antennal lobe (AL) are labeled. X-ray absorption coefficients were rendered from 20 cm$^{-1}$ (black) to 120 cm$^{-1}$ (white). (B) The same Drosophila brain visualized at 140 nm resolution. A total of 25 dataset volumes were fitted each other to compose the entire image. Phase shift values were rendered in arbitrary unit.

**Table 1.** Conditions for tomographic data collection and statistics of model building.

| Data collection conditions | |
|---|---|
| Beamline | BL47XU |
| X-ray energy (keV) | 8.0 |
| Pixel size (nm) [a] | 262 × 262 |
| Viewing field size (pixels) [a] | 2000 × 700 |
| Rotation/frame (degrees) | 0.10 |
| Exposure/frame (ms) | 300 |
| Frame/dataset | 1800 |
| Dataset collection time (min) | 16 |
| Number of observed pixels per dataset | 2520 × 10$^6$ |
| Number of datasets per sample | 4 |
| **Model building statistics** | |
| Size of coefficient map (μm) [b] | 220.1 × 327.5 × 314.4 |
| Number of coefficient map voxels | 1260 × 10$^6$ |
| Number of nodes | 251,970 |
| Number of traces | 15,179 |
| Total length of traces (mm) | 378.296 |

[a] Width × height

[b] Width × depth × height along the rotation axis

**Supplementary data legends**

**Supplementary video 1.** Structure of Drosophila brain. The structure rotates nearly around the dorsoventral axis, while anterior and posterior parts of the left hemisphere are gradually cut away to illustrate inner structures. X-ray absorption coefficients were rendered from 15 cm$^{-1}$ (black) to 100 cm$^{-1}$ (white) using the program VG Studio MAX (Volume Graphics, Germany) with the scatter HQ algorithm. Dorsal is to the top.

**Supplementary video 2.** Structure of optic lobe. First, the entire model rotates by 360° around the dorsoventral axis. Neuronal process groups are differentiated by color. Next, the view switches to the medulla structure (corresponding to Fig. 3B), then to the second optic chiasma and lobula (Fig. 3C and 3D), and finally to the lobula plate (Fig. 3E), while neuronal processes of the central brain are drawn in gray. Video frames were produced using the program MCTrace. Dorsal is to the top.

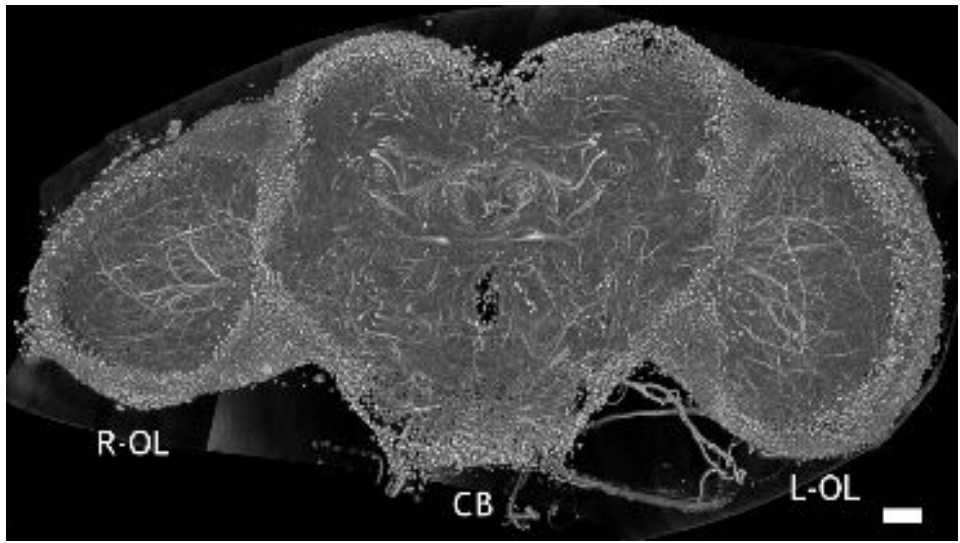
Fig 1A

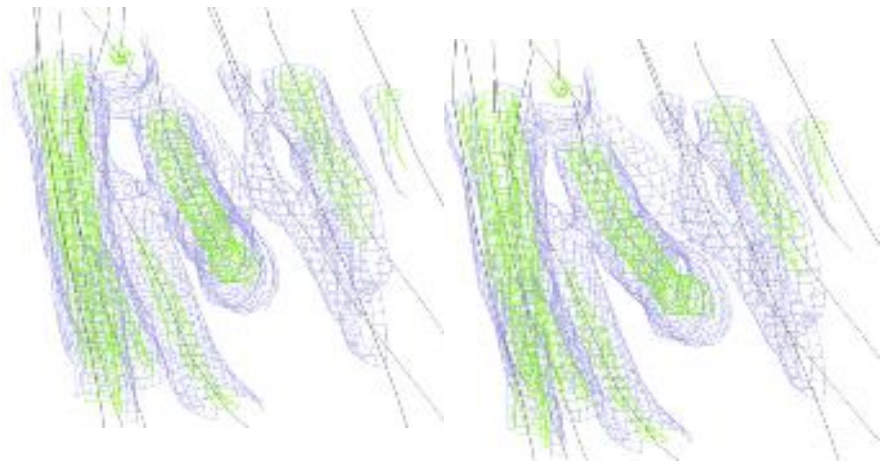
Fig 1B

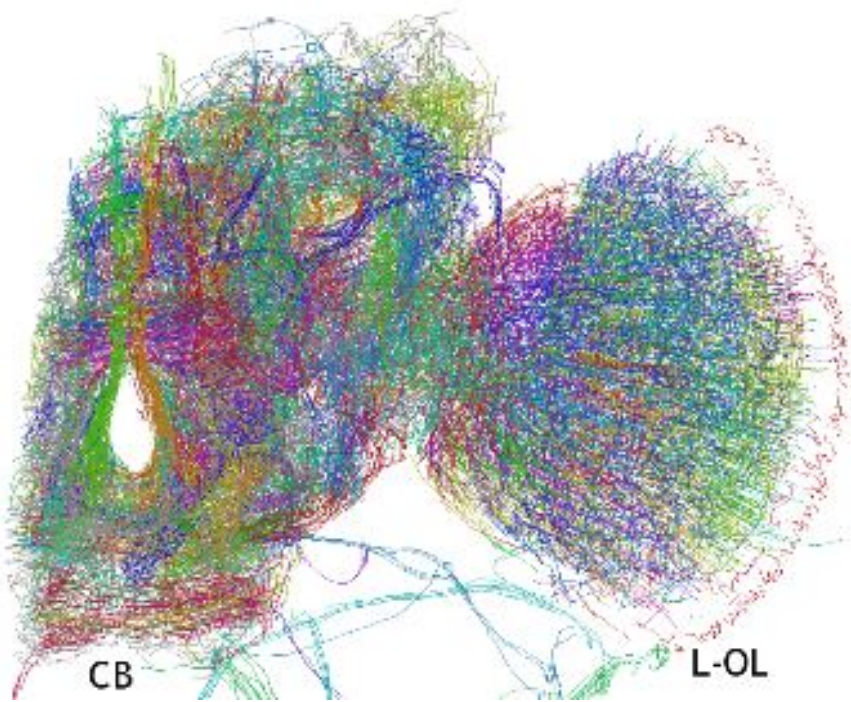
Fig 1C

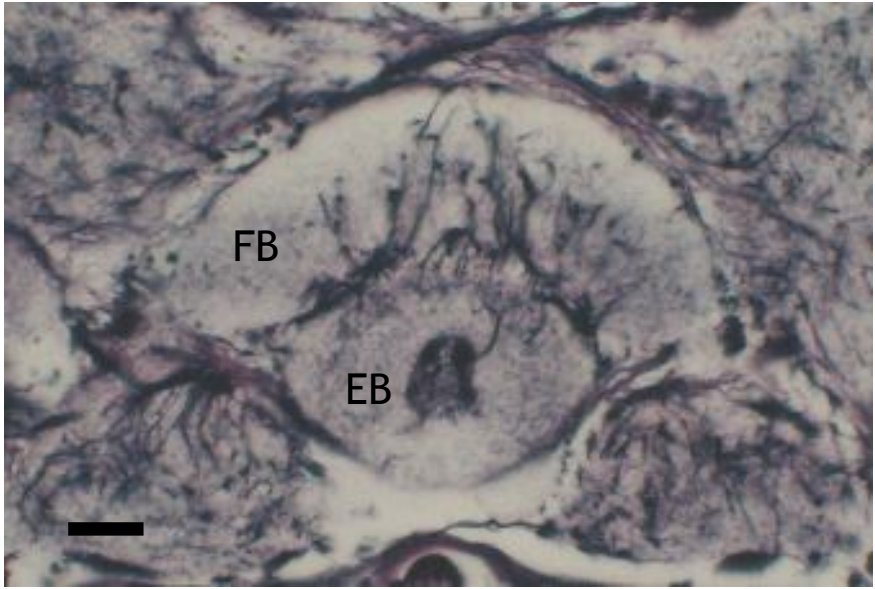

Fig 2A

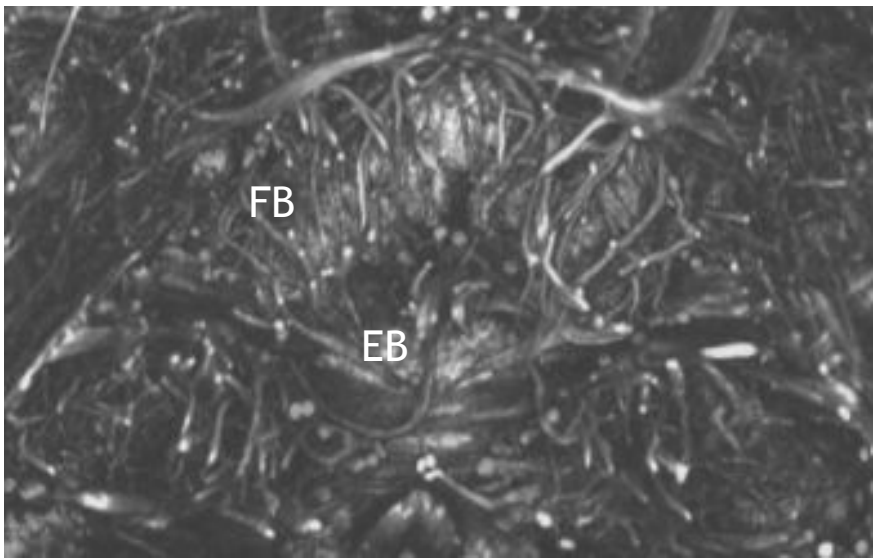

Fig 2B

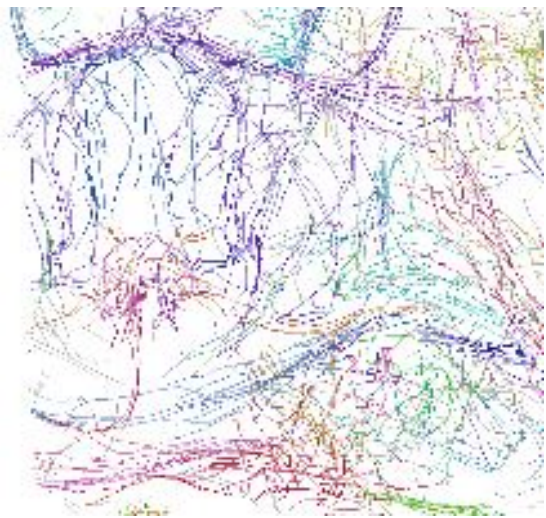

Fig 2C

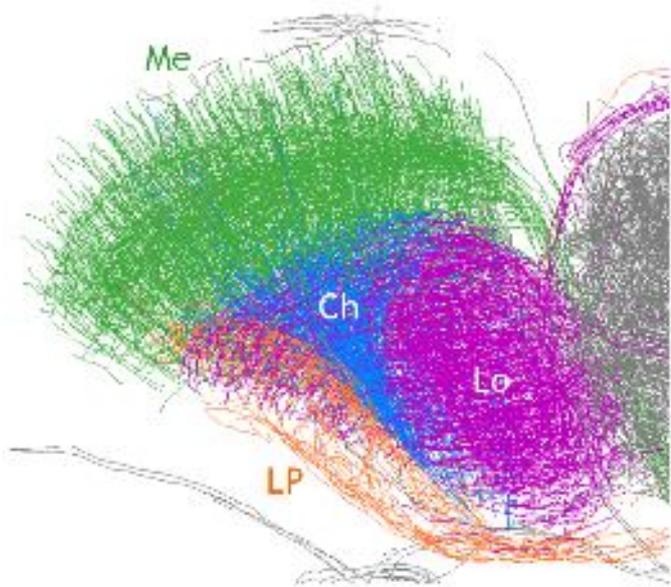

Fig 3A

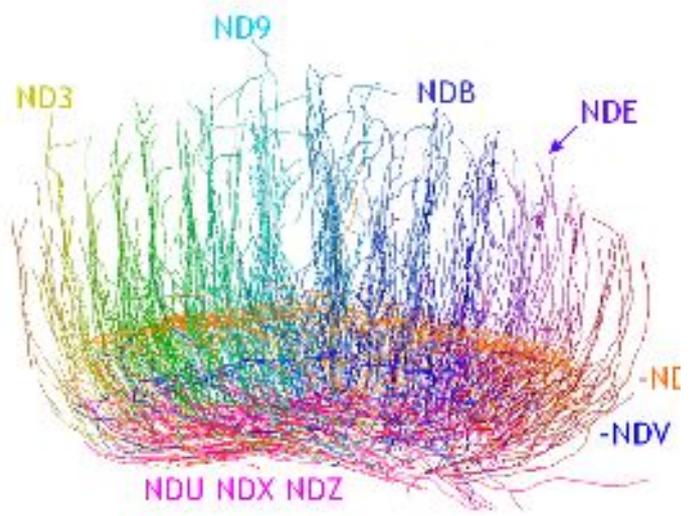

Fig 3D

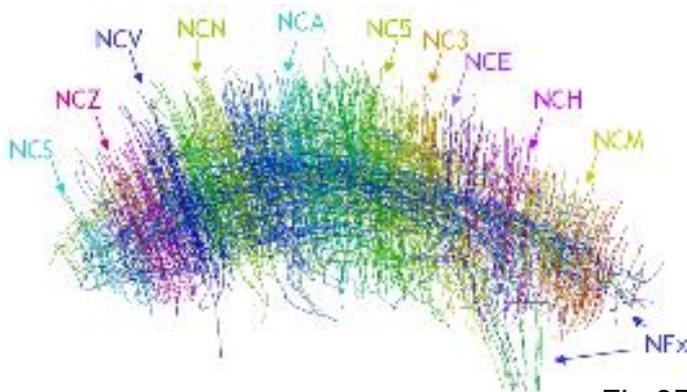

Fig 3B

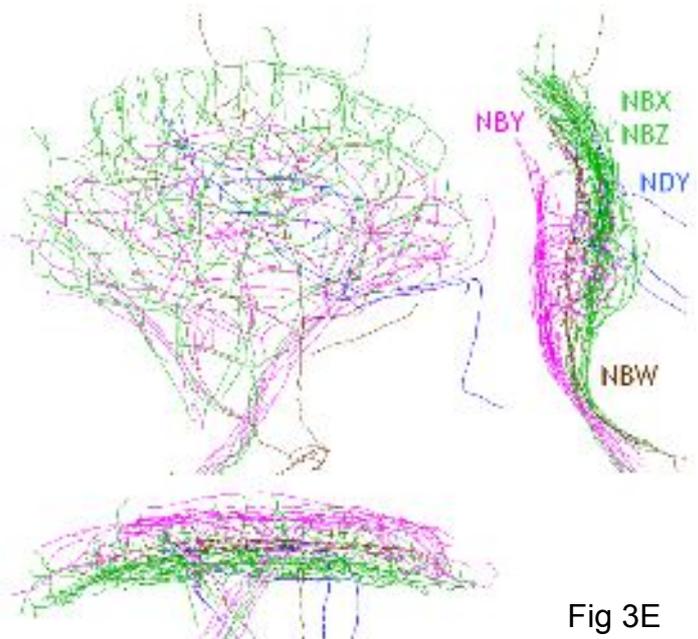

Fig 3E

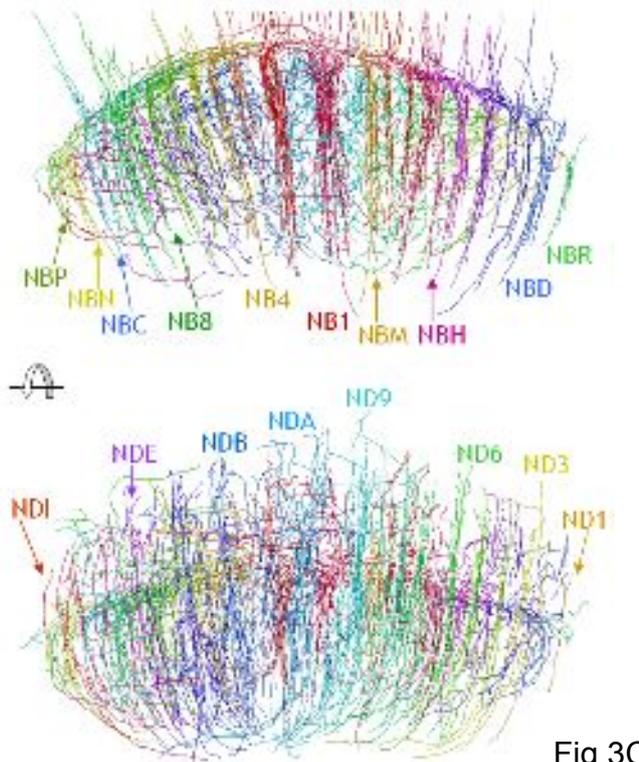

Fig 3C

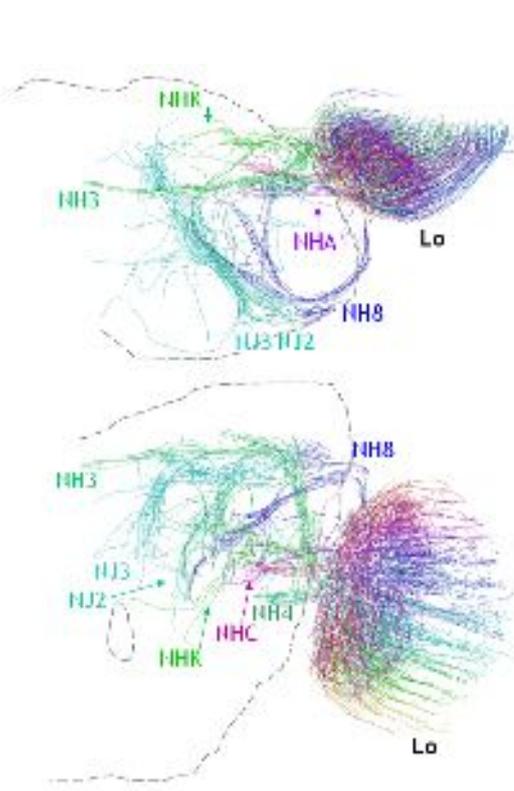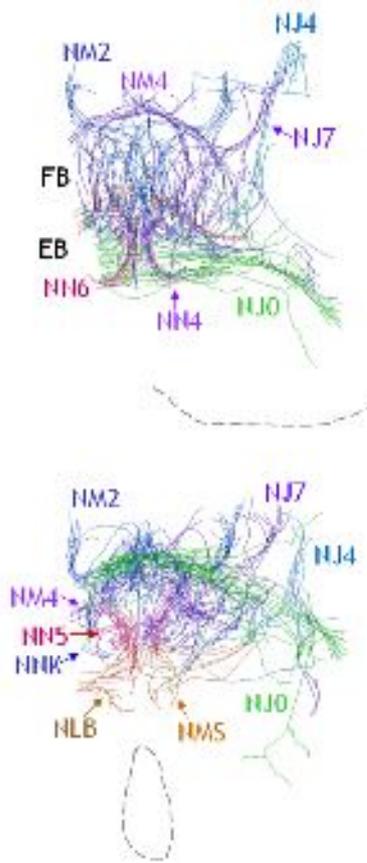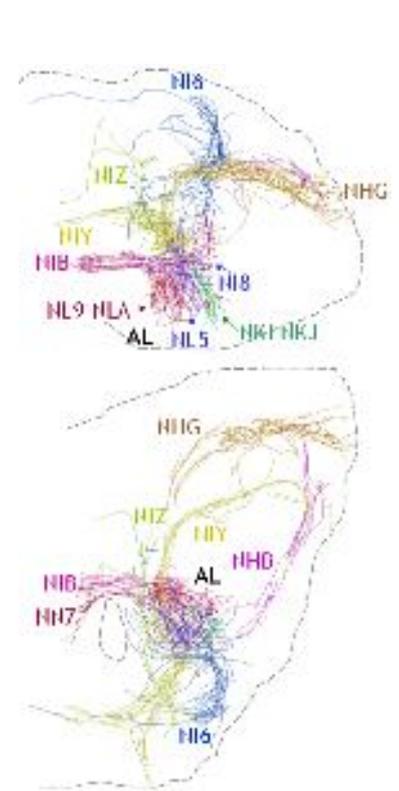

Fig 4A                         Fig 4B                         Fig 4C

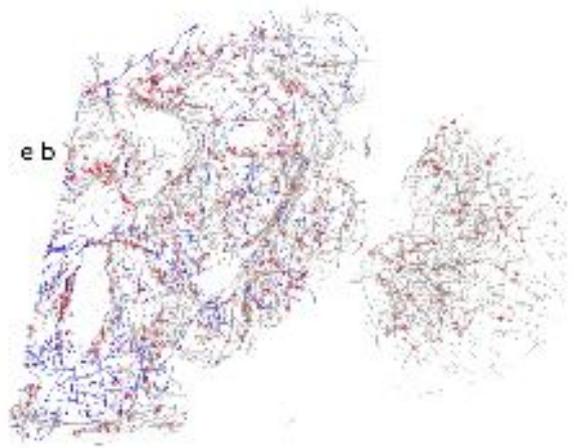

Fig 5A

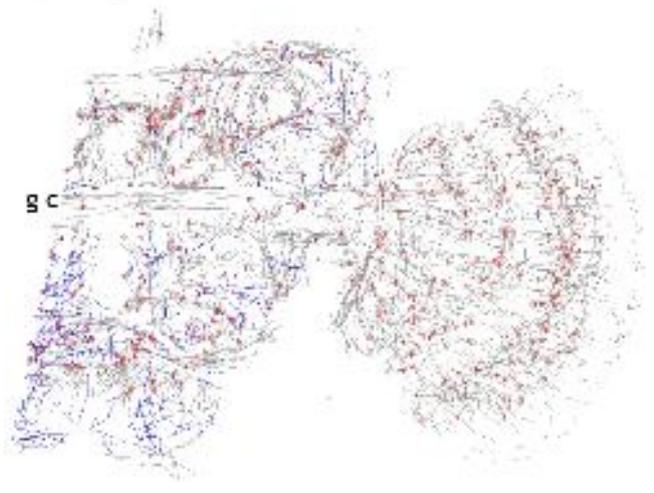

Fig 5B

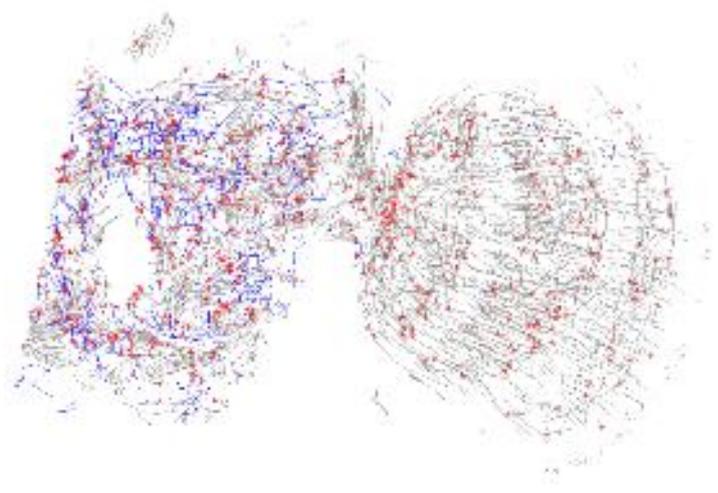

Fig 5C

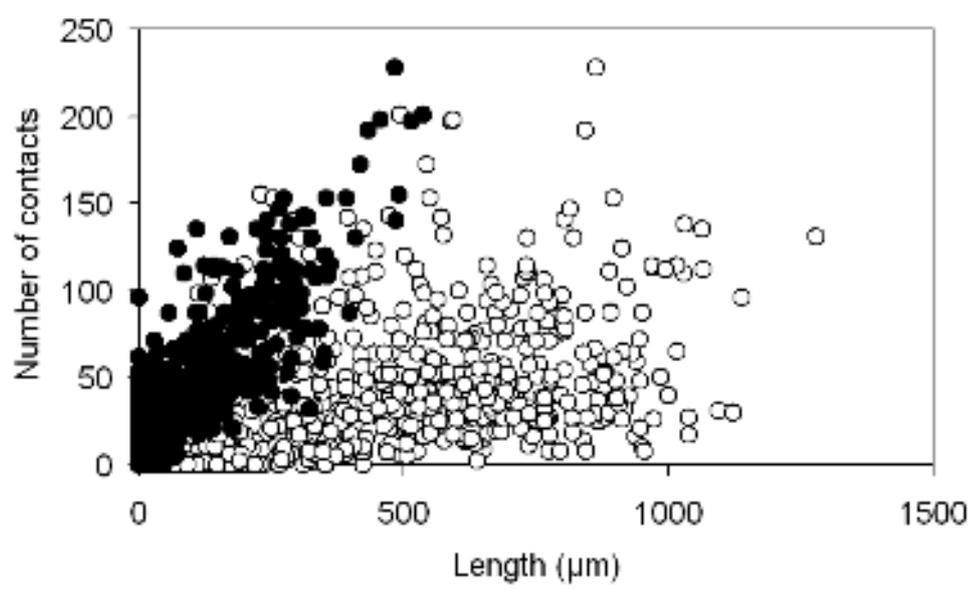

Fig 6

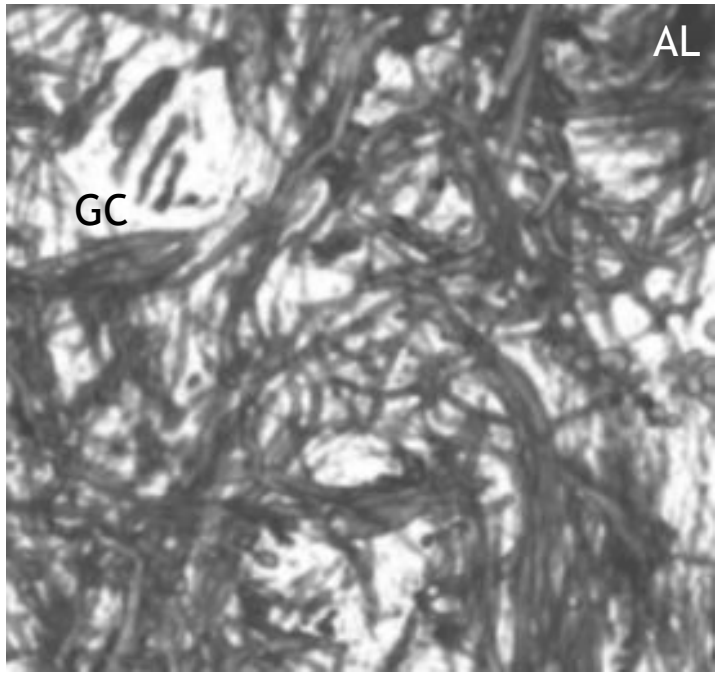

Fig 7A

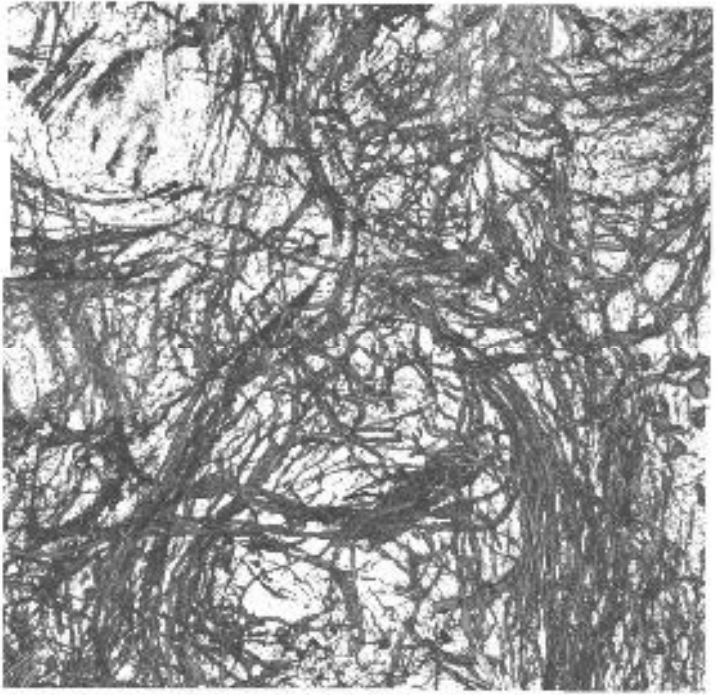

Fig 7B